\documentclass[twocolumn, pra, showpacs,superscriptaddress]{revtex4}%
\usepackage{amsfonts}
\usepackage{amsmath}
\usepackage{amssymb}
\usepackage{graphicx}
\usepackage{subfigure}
\usepackage{epstopdf}
\usepackage{multirow}%
\providecommand{\U}[1]{\protect\rule{.1in}{.1in}}
\begin{document}
\title{Coupled spin-vortex pair in dipolar spinor Bose-Einstein condensates}
\author{Tiantian Li}
\affiliation{Institute of Theoretical Physics, Shanxi University,
Taiyuan 030006, China}
\author{Su Yi}
\email{syi@itp.ac.cn}
\affiliation{State Key Laboratory of Theoretical Physics, Institute of Theoretical Physics,
Chinese Academy of Sciences, P.O. Box 2735, Beijing 100190, China}
\author{Yunbo Zhang}
\email{ybzhang@sxu.edu.cn}
\affiliation{Institute of Theoretical Physics, Shanxi University,
Taiyuan 030006, China}

\begin{abstract}
We investigate the ground-state and the magnetic properties of a dipolar spin-1 Bose-Einstein condensate trapped in a symmetric double-well potential. In particular, we focus on the spin-vortex states by assuming that each potential well is highly pancake-shaped. We show that the presence of the double-well potential gives rise to two different spin configurations for the spin-vortex pair states. We also study the response of the coupled spin-vortex pair to static transverse magnetic fields.
\end{abstract}


\maketitle

\section{Introduction}

Magnetic dipole-dipole interaction (MDDI) in spinor Bose-Einstein condensates has attracted much attention both experimentally~\cite{Stamper-Kurn08,Stamper-Kurn10,Laburthe-Tolra11,eto14} and theoretically \cite{Santos06,Ueda06,Yi06,SpontaneousPT,Machida,Yi08,Ho10} in recent years. Through coupling the spin and orbital angular momenta, the MDDI may gives rise to the Einstein-de Haas effect~\cite{Santos06,Ueda06} and spontaneous spin textures~\cite{Yi06,SpontaneousPT,Machida,Yi08,Ho10}. Indeed, it has been experimentally shown that the MDDI is responsible for the  decay of a helical spin structure in a spin-1 Rb condensate~\cite{Stamper-Kurn08,Stamper-Kurn10}, the spontaneous demagnetization of a spin-3 Cr condensate~\cite{Laburthe-Tolra11}, and the spin texture in a spin-2 Rb condensate~\cite{eto14}. Of particular interest, for sufficiently strong MDDI interaction, it was theoretically predicted that spin-vortex state (SVS) spontaneously forms in a spinor condensate~\cite{Yi06,SpontaneousPT}, which enhances the head-to-tail spin alignment and subsequently lowers the MDDI energy~\cite{Ueda}.

Spin vortices also appear in other contexts, most notably the magnetic vortex in ferromagnetic nanodisks resulting from the competition between the exchange energy and the magnetostatic energy~\cite{Hubert,nanomagnetism,C. L. Chien}. In general, a magnetic vortex consists of a planar spin curl which turns out of the plane near the vortex core. Other than the winding number (which is often of unit), magnetic vortices are characterized by their chirality (clockwise or counterclockwise curling direction) and polarity (up or down direction of the vortex core magnetization)~\cite{nanomagnetism,C. L. Chien,mv-0.5}, which gives rise to four independent combinations of chirality and polarity. Because of their topological nature, rich dynamic properties, and potential applications in information storage, magnetic vortices have been extensively studied over the past decades~\cite{mv-1,mv-0.5,mv0,mv1,mv2,mv3,mv4,mv5,mv6}. More interestingly, magnetic vortices in multilayer structures~\cite{multilay1,multilay2,multilay3,multilay4,multilay5,multilay6,multilay7,multilay8,multilay9,multilay10}, e.g., ferromagnetic/nonmagnetic/ferromagnetic (F/N/F), are of practical importance for memory devices. In addition, the interlayer coupling may lead to new physics, such as the unique vortex oscillations and meronlike state~\cite{Guslienko2005, Jun,Phatak}. 

Armed with the spontaneous SVS and the tunablity in both the atom-atom interactions and the geometries of the system, dipolar spinor condensates in a multiple-well potential provide an ideal platform for simulating the multilayer magnetic vortices. In particular, the recent successes in creating ultracold gases of polar molecules~\cite{KRb,LiCs,RbCs,RbCs2,RbCs3,NaK} offer an opportunities in studying the coupled SVS with higher winding number~\cite{Deng}.

By focusing on the spin vortex states, we investigate, in this paper, the ground-state structure of a dipolar spin-1 Bose-Einstein condensate trapped in a symmetric double-well potential, in analogous to the the F/N/F multilayer structure. The introduction of the double-well potential splits a spin vortex into a pair of spin vortices. In order to obtain a sufficiently strong interwell coupling, we assume that each potential well is highly oblate. In this case, the spin at the vortex core simply vanishes such that each spin vortex is only characterized by its chirality. We show that, by increasing the interwell barrier, a pair of spin vortices with same chirality converts into a pair with opposite chiralities, as the result of the interplay between the interwell MDDI and the tunneling splitting of the double-well potential. We also study the magnetization process of the spin-vortex pairs subjected to a transverse field. By orientating the spins, the transverse field destroys spin vortices when the condensate is fully polarized. Interestingly, the spin vortices in two potential wells disappear from the condensate sequentially, instead of being destroyed simultaneously.

This paper is organized as follows. In Sec.~\ref{form}, we introduce our model and specify parameters considered in this work. The numerical results about the ground-state structure are presented in Sec.~\ref{resu}. Finally, we conclude in Sec.~\ref{concl}.

\section{Formulation}\label{form}
We consider a condensate of $N$ spin $F=1$ atoms trapped in a double-well potential that is formed by imposing a Gaussian barrier upon a harmonic potential
\begin{align}
U\left( \mathbf{r}\right) =\frac{1}{2}M\left( \omega_{\perp}^{2}x^{2}+\omega_{\perp}^{2}y^{2}+\omega_{z}^{2}z^{2}\right) +Ae^{-z^{2}/(2\sigma
_{0}^{2})},\label{trap}
\end{align}
where $\omega_{\perp}$ is the radial trap frequency, $\omega_{z}$ is the axial trap frequency, and $A$ and $\sigma_{0}$ are the height and width of the Gaussian barrier, respectively. We note that it is important to distinguish the height of the Gaussian barrier $A$ and the height of the interwell barrier $\widetilde A$, as the latter is measured from the minima of the double-well potential. In fact, it is easy to show that $A$ and $\widetilde A$ are
related by the equation
\[
\widetilde A=A-M\omega_{z}^{2}\sigma_{0}^{2}\left( 1-\ln\frac{M\omega_{z}%
^{2}\sigma_{0}^{2}}{A}\right) .
\]

In the mean-field treatment, the condensate wave functions $\psi_{\alpha}({\mathbf{r}})$ ($\alpha=0,\pm1$) satisfy the coupled dynamical equations~\cite{Ho98,Machida98,Yi09}
\begin{align}
i\hbar\frac{\partial\psi_{\alpha}}{\partial t}=\big(T+U+c_{0}n\big)\psi
_{\alpha}+g_{F}\mu_{B}{\mathbf{B}}_{\mathrm{eff}}\cdot{\mathbf{F}}%
_{\alpha\beta}\psi_{\beta},\label{gpe}%
\end{align}
where $T=-\hbar^{2}\nabla^{2}/(2M)$ represents the kinetic energy with $M$ being the mass of the atom, $c_{0}=4\pi\hbar^{2}(a_{0}+2a_{2})/(3M)$ is the strength of the spin-independent collisional interaction with $a_{f}$
($f=0,2$) being the s-wave scattering length for two spin-1 atoms in the combined symmetric channel of total spin $f$, $n({\mathbf{r}})=\sum_{\alpha}|\psi_{\alpha}|^{2}$ is the total density, $g_{F}$ is the Land\'{e} $g$-factor of the atom, $\mu_{B}$ is the Bohr magneton, and ${\mathbf{F}}$ is the angular momentum operator. The effective magnetic field in Eq.~(\ref{gpe}) includes the contributions from the external magnetic field ${\mathbf{B}}_{\mathrm{ext}}$, the spin-exchange collisional interaction ${\mathbf{B}}_{\mathrm{col}}$, and the dipolar interaction ${\mathbf{B}}_{\mathrm{dip}}$, i.e.,
\begin{align}
{\mathbf{B}}_{\mathrm{eff}}({\mathbf{r}}) & ={\mathbf{B}}_{\mathrm{ext}%
}+{\mathbf{B}}_{\mathrm{col}}({\mathbf{r}})+{\mathbf{B}}_{\mathrm{dip}%
}({\mathbf{r}}),\label{beff}\\
{\mathbf{B}}_{\mathrm{col}}({\mathbf{r}}) & =\frac{c_{2}}{g_{F}\mu_{B}%
}{\mathbf{S}}({\mathbf{r}}), \nonumber\\
{\mathbf{B}}_{\mathrm{dip}}({\mathbf{r}}) & =\frac{c_{d}}{g_{F}\mu_{B}}%
\int\frac{d{\mathbf{r}}^{\prime}}{|{\mathbf{R}}|^{3}}\left[ {\mathbf{S}%
}({\mathbf{r}}^{\prime})-\frac{3\left[ {\mathbf{S}}({\mathbf{r}}^{\prime
})\cdot{\mathbf{R}}\right] {\mathbf{R}}}{|{\mathbf{R}}|^{2}}\right] ,\nonumber
\end{align}
where $c_{2}=4\pi\hbar^{2}(a_{2}-a_{0})/(3M)$, ${\mathbf{S}}({\mathbf{r}})=\sum_{\alpha\beta}\psi_{\alpha}^{*}{\mathbf{F}}_{\alpha\beta}\psi_{\beta}$ is the density of the spin, ${\mathbf{R}}={\mathbf{r}}-{\mathbf{r}}^{\prime}$, and $c_{d}=\mu_{0}\mu_{B}^{2}g_{F}^{2}/(4\pi)$ with $\mu_{0}$ being the vacuum magnetic permeability. 

For the numerical results presented in this work, we take $N=5\times10^{5}$ and $\omega_{\perp}=(2\pi)100\,$Hz. Since we are interested in the spin-vortex state, we adopt a pancake-shaped harmonic potential with the asymmetric parameter of the harmonic potential $\lambda=\omega_{z}/\omega_{\perp}=6$. Unless otherwise stated, the value of the barrier width is fixed at $\sigma_{0}=7\,\mu\mathrm{m}$. The $s$-wave scattering lengths are chosen as those of $^{87}$Rb atom, i.e., $a_{0}=5.40\,$nm and $a_{2}=5.32\,$nm. For ${}^{87}$Rb atom, the dipolar interaction strength $c_{d}$ is roughly $10$\% of the spin-exchange interaction strength $c_{2}$. In this work, we deliberately increase it to $c_{d}=|c_{2}|$ such that the condensate is a spin-vortex state in the absence of the barrier. We note that the purpose of choosing a rather large $c_{d}$ is to accelerate the convergence of the
numerical calculation and this choice does not qualitatively change the results presented below. In fact, it was verified that the qualitative results can be reproduced by solely increasing $N$ to $10^{7}$ with $c_{d}$ being that of the $^{87}$Rb atom. Finally, to ensure that the dipolar interaction energy is comparable to the linear Zeeman energy, the external magnetic field covered by our numerical simulations is below $100\,\mu{\rm G}$. Consequently, we neglect the quadratic Zeeman effect throughout this work~\cite{wxzhang}.

\section{Results}\label{resu} 
In this section, we investigate the ground-state properties of the dipolar spin-1 condensate in a double-well potential. To this aim, we treat $A$, $\sigma_{0}$, and ${\mathbf{B}}$ as control parameters and calculate the ground-state wave function by numerically evolving Eq.~(\ref{gpe}) in imaginary time. For convenience, we adopt the dimensionless units: $\hbar\omega_{\perp}$ for energy, $\ell_{\perp}=\sqrt{\hbar/(M\omega_{\perp})}$ for length, and $\ell_{\perp}^{-3/2}$ for the wave functions.

\subsection{Ground state without external magnetic field}

\begin{figure}[ptb]
\includegraphics[width=0.95\columnwidth]{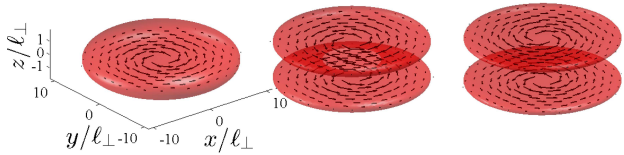} \centering
\caption{(color online). Ground state structures for barrier heights $A/(\hbar\omega_{\perp})=0$, $100$, and $300$, from left to right. The total densities are represented by isodensity surface plots for $n({\mathbf{r}
})=10^{14}\,\mathrm{cm}^{-3}$. The spin structures are shown by the vector plots on the $z=z_{\mathrm{min}}$ planes with $z_{\mathrm{min}}$ being the positions of the potential minima along the $z$ axis.}
\label{fdenspin}
\end{figure}

Before we present our results in the double-well potential, let us first recall the ground-state wave function in a pancake-shaped potential without the barrier, i.e., $A=0$. As shown in Ref.~\cite{Yi06}, the ground-state wave function can be expressed as
\begin{align}
\psi_{\alpha}({\mathbf{r}})=\sqrt{n_{\alpha}({\mathbf{r}})}e^{i\Theta_{\alpha
}({\mathbf{r}})},\label{denpha}%
\end{align}
where $n_{\alpha}=|\psi_{\alpha}|^{2}$ is the density of the $\alpha$-th spin component which is axially symmetric and
\begin{align}
\Theta_{\alpha}({\mathbf{r}})=w_{\alpha}\varphi+\vartheta_{\alpha
}\label{wfphase}%
\end{align}
is the corresponding phase with $w_{\alpha}$ being the winding numbers, $\varphi$ the azimuthal angle, and $\vartheta_{\alpha}$ the phase angles. Of particular interest, it was found that the densities and the winding numbers the wave function for a SVS satisfy the conditions~\cite{Yi06}:
\begin{align}
& n_{1}({\mathbf{r}})=n_{-1}({\mathbf{r}}),\label{density}\\
& (w_{1},w_{0},w_{-1})=(-1,0,1),\label{winding}%
\end{align}
which lead to the planar spin
\begin{align}
(S_{x},S_{y})=f({\mathbf{r}})\big(\cos(\varphi-\delta),\sin(\varphi
-\delta)\big),\label{espin}%
\end{align}
where
\begin{align}
f({\mathbf{r}})=2\sqrt{2n_{0}n_{1}}\cos\left[ \vartheta_{0}-(\vartheta
_{1}+\vartheta_{-1})/2\right] \label{funcf}%
\end{align}
and $\delta=(\vartheta_{1}-\vartheta_{-1})/2$. Moreover, the phase angles satisfy
\begin{align}
\vartheta_{0}-\frac{1}{2}(\vartheta_{1}+\vartheta_{-1})=0.\label{phase}%
\end{align}
Clearly, the conditions (\ref{density}), (\ref{winding}), and (\ref{phase}) give rise to the SVS shown in the left panel of Fig.~\ref{fdenspin}.

In a double-well potential, the gas is spatially split into two thin layers as the barrier height is increased. It is anticipated that the spin structure of the condensate should be determined by both intralayer and interlayer MDDI. In fact, as shown in the middle and right panels of Fig.~\ref{fdenspin}, we find that for $A<A^{*}=246\hbar\omega_{\perp}$, the spin vortices in the upper and lower wells have same chirality, which is referred to as parallel spin-vortex pair (PSVP). While for $A>A^{*}$, the spin vortices in the upper and lower layers have opposite chiralities, forming a antiparallel spin-vortex pair (ASVP). Intuitively, the PSVP to ASVP transition is induced by the interlayer MDDI which energetically favors the antiparallel spin alignment between different potential wells.

Now, let us examine the ground-state wave functions in detail. For convenience, we partition the wave function $\psi_{\alpha}$ into the lower part $\psi_{\alpha}^{(l)}$ (for $z<0$) and the upper part $\psi_{\alpha}^{(u)}$ (for $z\geq0$). Correspondingly, we may define $n_{\alpha}^{(i)}$, $\Theta_{\alpha}^{(i)}$, $w_{\alpha}^{(i)}$, and $\vartheta_{\alpha}^{(i)}$ ($i=l,u$) for the lower and upper parts. In the PSVP phase, it is found that the conditions (\ref{density}), (\ref{winding}) and (\ref{phase}) are satisfied by both the lower and upper parts of the wave function, which naturally leads to the PSVP. In the ASVP phase, although the conditions (\ref{density}) and (\ref{winding}) are still fulfilled, the phase-angle condition (\ref{phase}) only holds for the lower part of the wave function, i.e., $\vartheta_{0}^{(l)}-\left( \vartheta_{1}^{(l)}+\vartheta_{-1}^{(l)}\right) /2=0$. For the upper part, we find that $\vartheta_{0}^{(u)}-\left( \vartheta_{1}^{(u)}+\vartheta_{-1}^{(u)}\right) /2=\pi$. Consequently, Eq. (\ref{funcf}) gives rise to the ASVP. More specifically, our numerical results show that $\vartheta_{\pm1}^{(l)}=\vartheta_{\pm1}^{(u)}$ and $\Delta\Theta_{0}=\Theta_{0}^{(u)}-\Theta_{0}^{(l)}=\pi$ in the ASVP phase.

\begin{figure}[t]
\includegraphics[width=0.36\textwidth]{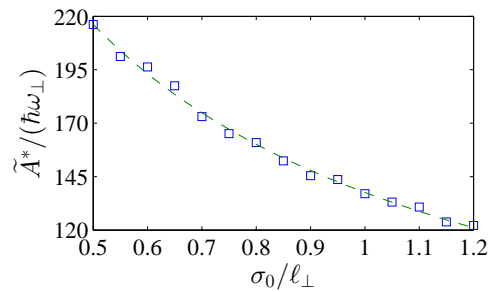}\centering
\caption{(color online). $\sigma_{0}$ dependence of $\tilde A^{*}$ obtained by numerical calculation (squares) and by fitting Eq. (\ref{fita}) (dashed line) with fitting parameters $\kappa_{1}=137.6$ and $\kappa_{2}=42.5$.}%
\label{avssig}%
\end{figure}

Physically, the PSVP to ASVP transition can be understood as follows. Let the real-valued functions $\phi^{(l)}$ and $\phi^{(u)}$ be, respectively, the ground-state wave functions of the lower and upper potential wells, the wave functions of the ground ($\psi_{0}^{(+)}$) and the first-excited ($\psi_{0}^{(-)}$) states of the spin $\alpha=0$ component in the double-well potential should be roughly $\psi_{0}^{(\pm)}\propto\phi^{(l)}\pm\phi^{(u)}$. The corresponding energies of these two states are separated by the tunneling splitting of the double-well potential. As one increases the barrier height, the tunneling splitting decreases. Eventually, at sufficiently large $A$, when the tunneling splitting cannot compensate the interlayer MDDI energy that favors the APSV state, $\psi_{0}^{(-)}$ becomes the ground-state wave function of the $\alpha=0$ spin component, for which we have exactly $\Delta\Theta_{0}=\pi$. It is worthwhile to mention that this $\pi$-phase difference between the lower and upper parts of $\psi_{0}$ in the ASVP phase is induced spontaneously by the MDDI.

The above argument can be further confirmed by considering the relation between $A^{*}$ and $\sigma_{0}$. In a double-well potential, the interlayer MDDI is proportional to $1/\sigma_{0}^{3}$ and the tunneling splitting is roughly proportional to $e^{-\widetilde A\sigma_{0}}$. By equating the tunneling splitting and the interlayer dipolar interaction, we obtain the critical barrier height of the double-well potential as
\begin{align}
\widetilde A^{*}=(\kappa_{1}+\kappa_{2}\ln\sigma_{0})\sigma_{0}^{-1}%
,\label{fita}%
\end{align}
where $\kappa_{1}$ and $\kappa_{2}$ are two constants. Figure~\ref{avssig} shows the $\sigma_{0}$ dependences of $\widetilde A^{*}$ obtained by the full numerical calculations and by fitting the Eq.~(\ref{fita}) with $\kappa
_{1}=137.6$ and $\kappa_{2}=42.5$, which demonstrates good agreement.

\begin{figure}[t]
\includegraphics[width=0.4\textwidth]{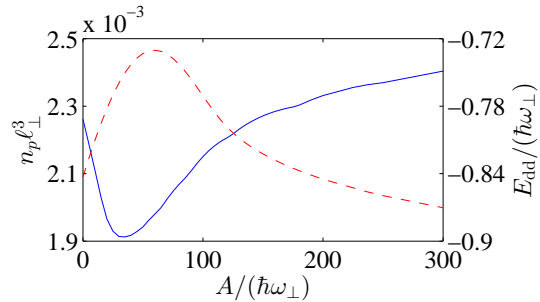}\centering
\caption{(color online). Barrier height dependence of the peak condensate density (solid line) and the MDDI energy (dashed line).}%
\label{peak}%
\end{figure}

It is also instructive to study the dependence of various physical quantities on the barrier height. In Fig.~\ref{peak}, we plot the peak condensate density $n_{p}=\mathrm{max}(n({\mathbf{r}}))$ and the MDDI energy per atom $E_{\mathrm{dip}}=g_{F}\mu_{B}N^{-1}\int d{\mathbf{r}}{\mathbf{S}}\cdot{\mathbf{B}}_{\mathrm{dip}}$ as functions of $A$. Depending on the value of $A$, those curves exhibit two distinct features: $n_{p}$ and $|E_{\mathrm{dip}}|$ decrease with $A$ in the small $A$ region; while in the large $A$ region, both $n_{p}$ and $|E_{\mathrm{dip}}|$ increase with growing $A$. To understand this, we note that, in Eq. (\ref{trap}), a double-well potential only forms when $A>M\omega_{z}^{2}\sigma_{0}^{2}$, for which, the effective axial trapping frequency of each potential well is $\omega_{\mathrm{eff}}=\omega_{z}\sqrt{2\ln [A/(M\omega_{z}^{2}\sigma_{0}^{2})]}$. Therefore, in the small $A$ region, the Gaussian barrier effectively flattens the harmonic potential along the $z$ direction. Consequently, $n_{p}$ and $E_{\mathrm{dip}}$ decrease with $A$. In the large $A$ region, the effective axial trapping frequency of each potential well increases with $A$. As a result, both the peak density and the dipolar interaction energy grows as one increases $A$.

\subsection{Ground state under a transverse magnetic field}

\begin{figure}[t]
\includegraphics[width=0.48\textwidth]{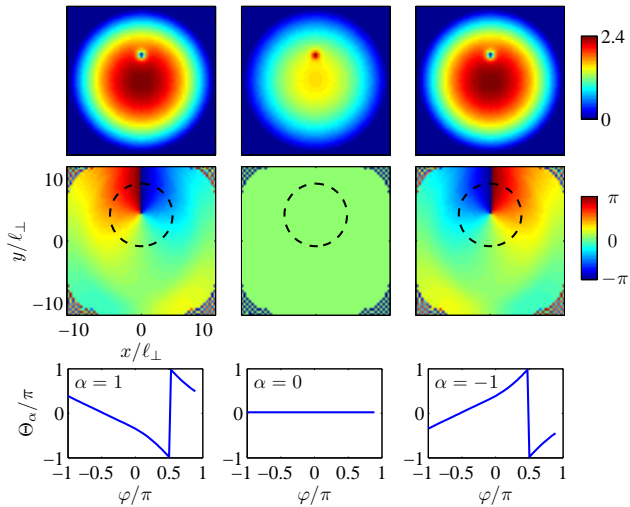}\centering
\caption{(color online). Ground state wave function on the $z=0$ plane for $A=0$ and $B_{x}=40\,\mu\mathrm{G}$. The upper panels show (from left to right) the densities (in units of $10^{14}\;{\rm cm}^{-3}$) for $\alpha=1$, $0$, and $-1$ components and the middle panels represent the corresponding phases. The lower panels show the phase $\Theta_{\alpha}$ as functions of the azimuthal angle along the dashed circles in the middle panels.}
\label{wave}%
\end{figure}

The magnetic properties of the spin-vortex state was previously investigated in Ref.~\cite{Yi06,Yi08}. It was shown that, depending on the direction of the magnetic field, the winding numbers of the condensate wave functions change from $(w_{1},w_{0},w_{-1})=(-1,0,1)$ to $(-2,-1,0)$ or $(0,1,2)$ under a sufficiently large longitudinal field. Such transition involves the changes of vorticity of each spin component and represents a first-order phase transition. It should be noted that, among the three spin states, the majorly populated spin component is always vortex free and the population in the vortex states becomes negligibly small under a strong magnetic field. It was also briefly mentioned in Ref.~\cite{Yi08} that, under a transverse magnetic field, the core of the spin vortex moves away from the center of the trap along the direction perpendicular to the external field and eventually disappears.

\begin{figure}[t]
\includegraphics[width=0.35\textwidth]{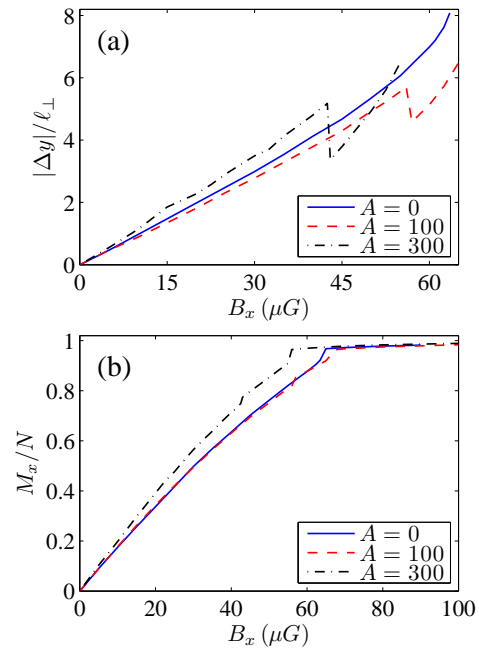}\centering
\caption{(color online). Transverse field dependence of the displacement of the spin vortex core $\Delta y$ (a) and the magnetization $M_{x}$ (b) for various barrier heights.}%
\label{svorbx}%
\end{figure}

Now, we present a detailed investigation for the spin-vortex states subjected to a transverse field, say ${\mathbf{B}}_{\mathrm{ext}}=B_{x}\hat{\mathbf{x}}$. First we note that the transverse magnetic field $B_{x}$ forces the spin to polarize along the $x$ direction. Therefore, in a spin vortex state, the population of atoms with spin parallel to the magnetic field increases. Consequently, the vortex core shifts along the $y$ direction such that the number of atoms with spin antiparallel to the external field is decreased. Figure~\ref{wave} shows the densities and phases of the condensate wave functions for $A/(\hbar\omega_{\perp})=0$ and $B_{x}=40\,\mu\mathrm{G}$. As can be seen, except for the displacements of the vortex cores along the $y$ axis, the vorticity of each spin component remains unchanged. A close inspection of the phases of the condensate wave functions (the third row of Fig.~\ref{wave}) reveals that $\Theta_{\pm1}$ are no longer the linear functions of the azimuthal angle $\varphi$ as compared to those in the absence of external field. Quantitatively, the solid line of Fig.~\ref{svorbx}(a) shows the $B_{x}$ dependence of the displacement of vortex core, $\Delta y$. Clearly, $\Delta y$ increases monotonically with $B_{x}$ and the spin vortex disappears at a critical magnetic field $B_{x}^{*}=63.5\,\mu\mathrm{G}$. The solid line of Fig.~\ref{svorbx}(b) plots the magnetization $M_{x}=\int d{\mathbf{r}} S_{x}({\mathbf{r}})$ as a function of $B_{x}$. As can be seen, at $B_{x}^{*}$ where the spin-vortex disappears, the $M_{x}$ changes abruptly, signaling a transition from the SVS to the polarized (along the $x$ axis) vortex-free state.

We remark that the phase transitions induced by the longitudinal field and by the transverse field. In the former case, vorticity always presents in the condensate wave function; while in the latter, vortices may disappear completely. Under a large transverse field, the population in individual spin state is always comparable, therefore a vortex state in any spin component is energetically unfavorable due to the extra kinetic energy associated with the vortex. While in a longitudinal field, the occupation number in the vortex state may become negligibly small, whose contribution to the total energy is ignorable.

\begin{figure}[t]
\includegraphics[width=0.45\textwidth]{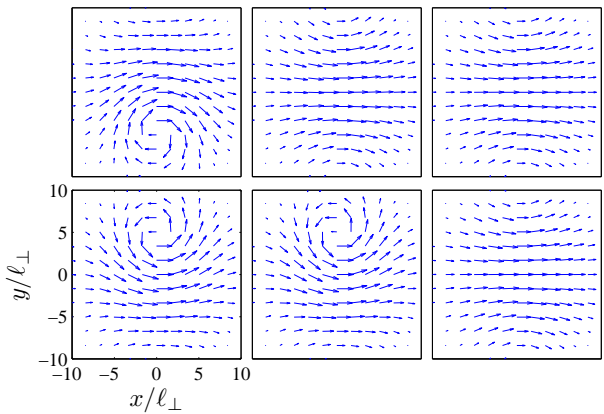}\centering
\caption{(color online). Spin structures in the upper (first row) and lower (second row) potential wells under the magnetic fields $B_{x}=40$ (left column), $50$ (middle column), and $70\,\mu\mathrm{G}$ (right column) for $A/(\hbar\omega_{\perp})=300$.}%
\label{spin}%
\end{figure}

In a double-well potential, it is apparent that the spin vortices in the upper and lower layers should move along the same direction for the PSVP state and along the opposite directions for the ASVP state. Interestingly, we find that, instead of vanishing simultaneously under a sufficiently large magnetic field, two spin vortices disappear from the condensate sequentially at the magnetic field strength $B_{x}^{*}$ and $B_{x}^{**}$. As an example, Fig.~\ref{spin} shows the typical spin structures corresponding to three stage of the magnetization process for an ASVP state. As can be seen, the number of spin vortices decreases from $2$ to $0$ with the growing magnetic field.

To gain more insight into these transitions, we also plot the $B_{x}$ dependence of $\Delta y$ for the vortex core in one of the potential wells for the PSVP and ASVP states in Fig.~\ref{svorbx}(a). Since $|\Delta y|$'s of the upper and lower spin vortices are always equal when $B_{x}<B_{x}^{*}$, we only plot $\Delta y$ corresponding to the vortex that disappears at $B_{x}^{**}$. Surprisingly, when the first vortex vanishes at $B_{x}^{*}$, $|\Delta y|$ of the second vortex drops, indicating that this vortex moves towards the center of the condensate. Physically, by getting closer to the condensate center, the layer with spin vortex contains more spins that are antiparallel to the spins in the other layer such that interlayer MDDI is lowered. In Fig.~\ref{svorbx}(b), we plot the magnetization $M_{x}$ as a function of $B_{x}$ for the PSVP and ASVP states. As can be seen, the transitions at $B_{x}^{*}$ can also be identified from the $M_{x}(B_{x})$ curves.

\section{Conclusions}

\label{concl} To conclude, we have investigated the ground-state structure of a dipolar spin-1 Bose-Einstein condensate trapped in a double-well potential by focusing on the spin-vortex states. We show that, in the absence of external magnetic field, the interplay of the tunneling splitting of the double-well potential and the interwell MDDI gives rise to the transition from the PSVP state to the ASVP state. Subsequently, we study the magnetization process of the spin-vortex state subjected to a transverse magnetic field. It has been shown that a spin vortex moves away from the condensate and disappears completely under sufficiently large transverse field. In particular, we find that a pair of spin vortices in a double-well potential vanish from the condensate sequentially.

\begin{acknowledgments}
This work was supported by the NSFC (Grant Nos. 11234008 and 11474189) and by the National 973 Program (Grant Nos. 2011CB921601 and 2012CB922104). SY acknowledge the support from NSFC (Grants Nos. 11434011 and 11421063).
\end{acknowledgments}

\end{document}